\documentclass[aps,prb,twocolumn,showpacs,preprintnumbers,amsmath,amssymb,superscriptaddress,longbibliography,nofootinbib]{revtex4-2}
\usepackage{hyperref}
\hypersetup{colorlinks=true, citecolor=blue, linkcolor=blue,urlcolor=cyan}
\usepackage{color}
\usepackage{graphicx}
\usepackage{dcolumn}
\usepackage{braket}
\usepackage{amsthm}
\usepackage[caption=false]{subfig}
\usepackage{enumerate}

\definecolor{darkgreen}{cmyk}{0.9,0,0.9,0}

\newcommand{\ex}[1]{\mathrm{e}^{#1}}
\newcommand{\fr}{\frac}
\newcommand{\pa}[1]{\left(#1 \right)}

\newcommand{\abs}[1]{\left|#1\right|}

\def\del{{\partial}}

\newcommand{\Disp}{{\rm D}}

\makeatletter
\let\cat@comma@active\@empty
\makeatother

\begin{document}
\preprint{CALT 2025-041}
\preprint{IPMU 25-0054}
\preprint{KYUSHU-HET-333}
\preprint{RIKEN-iTHEMS-Report-25}
\title{Continuous Family of Conformal Field Theories\\
and Exactly Marginal Operators}

\author{Shota Komatsu}\email[]{shota.komatsu@cern.ch}
\affiliation{\it Department of Theoretical Physics, CERN, 1211 Meyrin, Switzerland.}

\author{Yuya Kusuki}\email[]{kusuki.yuya@phys.kyushu-u.ac.jp}
\affiliation{\it Institute for Advanced Study, 
Kyushu University, Fukuoka 819-0395, Japan.}
\affiliation{\it Department of Physics, 
Kyushu University, Fukuoka 819-0395, Japan.}
\affiliation{\it RIKEN Interdisciplinary Theoretical and Mathematical Sciences (iTHEMS),
Wako, Saitama 351-0198, Japan.}

\author{Marco Meineri}\email[]{marco.meineri@unito.it}
\affiliation{\it Dipartimento di Fisica, Universit\`a di Torino and INFN -- Sezione di Torino, \\
Via P. Giuria 1, Torino 10125, Italy.}

\author{Hirosi Ooguri}\email[]{ooguri@caltech.edu}
\affiliation{\it Walter Burke Institute for Theoretical Physics \rm \& \it Leinweber Forum for Theoretical Physics, \\ California Institute of Technology, Pasadena, CA 91125, USA}
\affiliation{\it Kavli Institute for the Physics and Mathematics of the Universe \rm (WPI), \it University of Tokyo, Kashiwa 277-8583, Japan}

\begin{abstract}
Does a conformal manifold imply the existence of exactly marginal operators? We answer this question affirmatively under the assumption that there exists a conformal interface with certain properties connecting nearby CFTs. We show that the exactly marginal operator that connects the CFTs can be reconstructed from the interface displacement operator. Our construction is model-independent and based on the general principles of conformal symmetry.
\end{abstract}
\maketitle
\section{Introduction and Summary}

If exactly marginal operators exist in a conformal field theory (CFT), they generate a conformal manifold, which parameterizes a continuous family of CFTs \cite{Zamolodchikov1986}. One might naturally ask if the converse also holds: Does a conformal manifold require the existence of exactly marginal operators? In this letter, we address this question by deriving the existence of exactly marginal operators under certain assumptions.

The existence of exactly marginal operators has important implications for the Swampland program, particularly for the Distance Conjecture \cite{Ooguri:2006in}. The earliest distance conjecture may have been stated by Albert Einstein in his autobiographical notes, in which he claimed that there are no free parameters in the basic equations of physics. This was reformulated  as Conjecture 0 in \cite{Ooguri:2006in} as “the moduli space of a consistent quantum gravity is parameterized by inequivalent expectation values of massless scalar
fields.” The existence of exactly marginal operators for a conformal manifold allows us to prove Conjecture 0 for gravitational theories in anti-de Sitter (AdS) space  as follows \cite{Ooguri:2022}:
If there is a continuous family of gravitational theories in AdS, then there must be a corresponding family of CFTs. 
If we can show that each CFT parameter is associated with an exactly marginal operator, the AdS/CFT dictionary would tell us that the exactly marginal operator corresponds to a 
massless scalar field with a flat potential, and the asymptotic behavior of this field determines the parameters.
Therefore, every parameter in quantum gravity in AdS is related to a value of a dynamical field, as claimed in Conjecture 0.\footnote{There has been an interesting development in interpreting the conjecture in AdS \cite{Banks:2025nfe,Sen:2025bmj,Sen:2025ljz,Sen:2025oeq}.
}

We note an interesting analogy with the absence of global symmetry in quantum gravity in AdS, which was proven in \cite{Harlow:2018jwu,Harlow:2018tng} using the splittability of symmetry generators \cite{Buchholz:1985ii} in CFT. For continuous symmetry, the splittability is the consequence of the well-known fact that the symmetry generators are expressed as exponentiated integrals of the Noether currents over a Cauchy surface. If we regard a continuous deformation of CFT as $(-1)$-form global symmetry, the existence of the corresponding exactly marginal operator is the splittability of the symmetry. In this way, the absence of free parameters and the absence of global symmetry in gravitational theories in AdS are unified and understood in a single framework as consequences of the splittability of symmetry generators.

\begin{figure}[t]
 \begin{center}
  \includegraphics[width=6cm,clip]{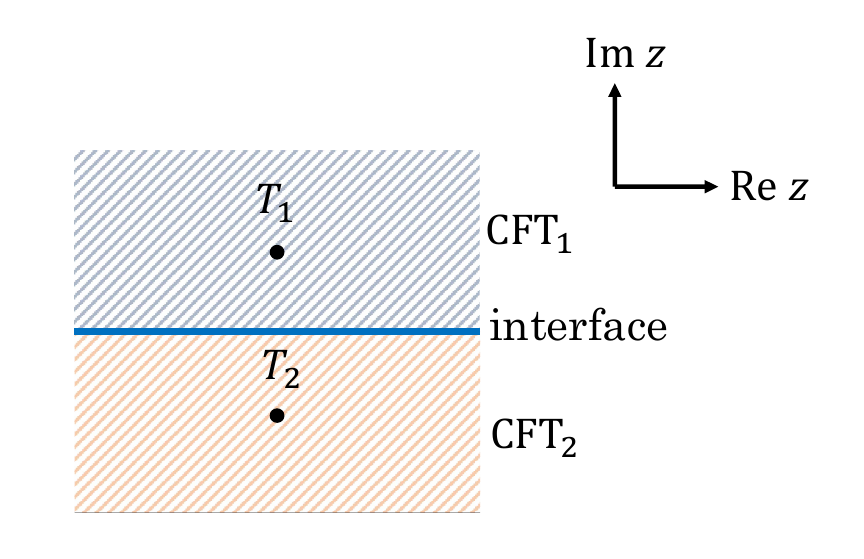}
 \end{center}
 \caption{Interface CFT. We denote the stress tensors of the CFTs on the upper and bottom sides of the interface by $T_1$ and $T_2$, respectively.
}
 \label{fig:interface}
\end{figure}

In what follows, we show that exactly marginal operators exist, under the assumption of the existence of a conformal interface between nearby CFTs that reduces to the identity when the CFTs coincide. The argument is constructive and model-independent.
We focus on two-dimensional CFTs for simplicity, but the proof can be generalized to higher dimensions and will be reported in \cite{Komatsu}.

\begin{figure}[t]
 \begin{center}
  \includegraphics[width=8.5cm,clip]{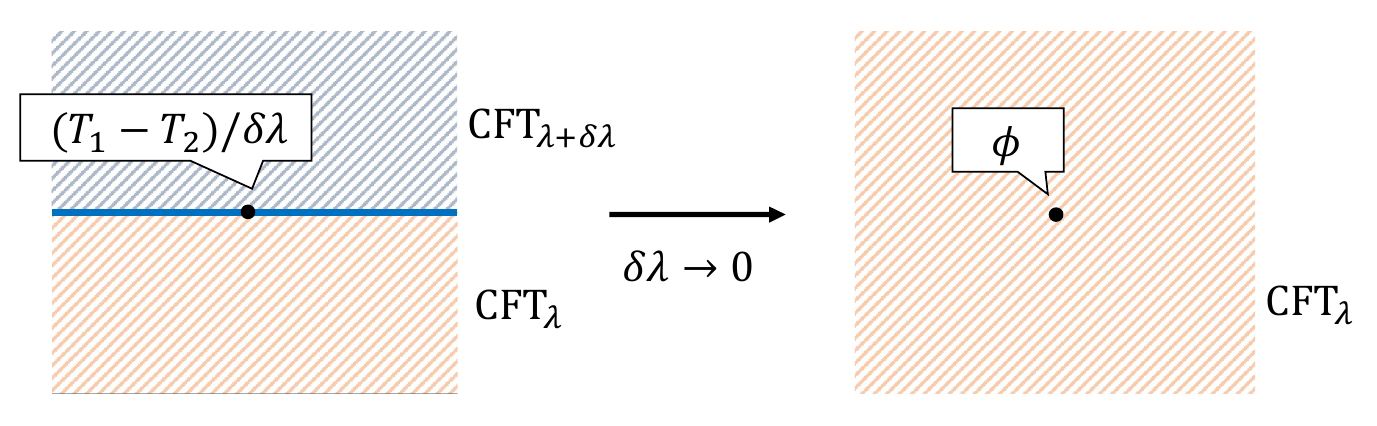}
 \end{center}
 \caption{A rough sketch of our construction of the exactly marginal operator $\phi$.
 We parameterize the conformal manifold by $\lambda$.
 As $\delta\lambda \to 0$, the two CFTs on both sides coincide,  
the interface vanishes, and what remains is the bulk local operator.
We show that this local operator is actually the exactly marginal operator.
 }
 \label{fig:construction}
\end{figure}

To motivate our proposal, consider a theory with an exactly marginal operator $\phi$, which we can use to deform the CFT in the upper half-plane as (See FIG. \ref{fig:interface} for our choice of convention)
\begin{equation}\label{eq:S}
S_0 \rightarrow S_0 + \delta\lambda \int_{\mathrm{Im}z>0} d^2 z \ \phi(z,\bar z),
\end{equation}
where $S_0$ is the original action of the CFT. At the leading order in $\delta \lambda$, this defines a conformal interface satisfying the gluing condition at ${\rm Im} \, z=0$ \cite{Oshikawa1996,Oshikawa1997,Bachas2002},
\begin{equation}\label{eq:bdy}
T_1-\overline{T}_1=T_2-\overline{T}_2,
\end{equation}
where $T_{1,2}$ ($\overline{T}_{1,2}$) are the holomorphic (anti-holomorphic) stress tensors on the upper and bottom half planes. Furthermore, it follows from the Ward identity that 
\begin{equation}\label{eq:Tphi}
T_1(x+i 0)-T_2(x-i 0)=\pi \delta \lambda \phi (x)+O(\delta \lambda^2)\,,
\end{equation}
where we have written $z=x+ iy$. Now we consider the displacement operator associated with this conformal interface \cite{Billo:2016cpy},
\begin{equation}
\mathrm{D} \equiv T_1 + \overline{T}_1 - T_2 - \overline{T}_2=2(T_1-T_2).
\end{equation}
The two-point function of the displacement operator has the following form,
\begin{equation}
\braket{\mathrm{D}(z)\mathrm{D}(w) } = \fr{\mathrm{N}_{\mathrm{D}}}{(z-w)^4}\,,
\end{equation}
where $\mathrm{N}_{\mathrm{D}}$ can be computed using \eqref{eq:bdy} and \eqref{eq:Tphi} as \cite{Brunner2016}
\begin{equation}
\mathrm{N}_{\mathrm{D}} = (2\pi \delta \lambda)^2 + O(\delta \lambda^3).
\end{equation}
This shows that one can extract the marginal operator $\phi(x)$ from (see FIG. \ref{fig:construction})
\begin{equation}\label{eq:maineq}
 \phi = \lim_{\delta \lambda \to 0} \hat{\mathrm{D}},
\end{equation}
where $\hat{\mathrm{D}}$ is the normalized displacement operator, 
\begin{align}
\hat{\mathrm{D}} \equiv \mathrm{D}/\sqrt{\mathrm{N}_{\mathrm{D}}}.
\end{align}

In the rest of this letter, we show that one can reverse this logic; assuming the existence of a conformal interface between nearby CFTs that reduces to the identity when the CFTs coincide, we provide a simple proof for the existence of an exactly marginal operator. It is desirable to prove the result without assuming the existence of such an interface. We are actively investigating this and hope to present details in an extended version forthcoming \cite{Komatsu}.

\section{Displacement Operator as Exactly Marginal Operator}
We define a conformal manifold as a collection of CFT data which depend continuously on a set of parameters. For every choice of parameters, the stress tensor exists and generates two copies of the Virasoro algebra. Our proof further assumes that
\begin{enumerate}
\item For any pair of CFTs that are close to each other on the conformal manifold, there is a conformal interface that becomes trivial ({\it i.e.} the identity interface) when the two CFTs coincide.
\item Correlators involving unit normalized defect operators have finite limits as the interface disappears (\emph{smoothness condition}).\footnote{While the Cauchy-Schwarz inequality implies that correlators of a single interface operator and any number of bulk operators are bounded, it does not imply that the limit exists. Furthermore, the norm of multiple defect insertions is not automatically bounded. It might be possible to improve this by considering multiple interfaces, but we do not pursue this avenue here.} 
\end{enumerate}
The consequence of these two assumptions is that defect operators become local operators of the CFT in the limit, unless they decouple (see FIG. \ref{fig:construction}).\footnote{If the interface in the limit is not trivial but topological and non-invertible, defect operators may become operators with a topological line attached.} This is true, in particular, for the (normalized) displacement operator $\hat{\mathrm{D}}$, which exists on every conformal interface. For the reasons explained later, we also make the additional technical assumption that 
\begin{enumerate}
\item[3.]The bulk CFT has no operators with conformal weights of $(1/2, 1/2)$ that are neutral under all the global symmetries preserved by the interface. 
\end{enumerate}
Note that this third assumption excludes Liouville theory from our discussion\footnote{For arbitrary values of the coupling constant, Liouville theory has a symmetry-neutral $(1/2,1/2)$ operator, hence violating our assumption 3. This operator may not correspond to a normalizable state in Liouville theory, but our current argument is insensitive to such a normalizability issue.}.

On the other hand, we do not need to assume that the central charges on both sides ($c_{L,R}$) are the same. Furthermore, chiral and antichiral central charges are allowed to differ, while the condition $c_1-\bar{c}_1=c_2-\bar{c}_2$ is necessary for the existence of a conformal interface \cite{Billo:2016cpy}. In other words, the central charge can vary on the conformal manifold, but $c-\bar{c}$ cannot.\footnote{
From modular $T$-invariance,
\begin{equation}
  h - \bar h - \frac{c - \bar c}{24} \in \mathbb{Z},
\end{equation}
while the single-valuedness of correlation functions implies the quantization of spin, $ h - \bar h \in \mathbb{Z}/2$.
Combining these relations, we conclude that $c - \bar c$ must be quantized.
This, in turn, means that $c - \bar c$ is invariant under continuous deformations.
}

Our proof proceeds by showing the following three facts:
\begin{enumerate}
  \item $\lim_{{\rm N}_{\rm D} \to 0}$$\hat{\mathrm{D}}$ is an interface primary operator of conformal dimension 2.
  \item $\lim_{{\rm N}_{\rm D} \to 0} \hat{\mathrm{D}}$ 
  is in fact a bulk primary operator of conformal dimension $(1,1)$.
  \item The three-point function of $\lim_{{\rm N}_{\rm D} \to 0} \hat{\mathrm{D}}$ vanishes.
\end{enumerate}

Here, an interface primary operator refers to a primary operator localized on the interface,
associated with a highest weight state under the single Virasoro symmetry that remains unbroken in the interface CFT. Interface quasiprimaries are defined analogously with respect to the $sl(2,\mathbb{R})$ algebra preserved by the defect.
Note that, unlike bulk scaling operators, which are labeled by conformal weights $(h,\bar{h})$ under the bulk symmetry $\mathrm{Vir} \otimes \overline{\mathrm{Vir}}$,  
interface scaling operators are characterized by a single eigenvalue corresponding to the remaining Virasoro symmetry.

Let us show that the displacement operator $\hat{\mathrm{D}}$
becomes an interface primary of weight $2$ in the limit where the two CFTs coincide. When evaluated on the interface, the stress tensors reorganize in two interface primaries ($W$, $\bar{W}$) and a quasiprimary ($T_{tot}=T_1+\overline{T}_2$), see \cite{Quella2007}.
Notice that the gluing condition implies $T_{tot}=\overline{T}_{tot}$. In particular, the displacement operator ${\rm D}$ can be shown to be an interface quasiprimary.
Indeed, one can check that 
\begin{align}\label{eq:OPE}
&\pa{T_1(z)+\overline{T}_2(z)}\hat{\mathrm{D}}(w)=\frac{(c_1-c_2)/\sqrt{\mathrm{N}_\mathrm{D}}}{(z-w)^4}\\
& +\pa{ \frac{2}{(z-w)^2} + \frac{1}{z-w}\partial }
\hat{\mathrm{D}}(w)+O(1),\nonumber
\end{align}
where $z$ and $w$ lie on the real axis,
and $O(1)$ denotes terms that are non-singular at $z = w$. Equation \eqref{eq:OPE} is obtained by making use of the interface boundary condition \eqref{eq:bdy}, as well as of the fact that $T_1(z)\overline{T}_2(w)$ and $\overline{T}_1(z)T_2(w)$ have no singularities at $z = w$.
This absence of singularity follows from the cluster decomposition principle \cite{Billo:2016cpy}.

In the limit ${\rm N}_{\rm D}\to 0$, the singular part of \eqref{eq:OPE} vanishes, thanks to the Averaged Null Energy conditions, which enforce the bound $\mathrm{N}_\mathrm{D} \geq 2 \abs{c_1-c_2}$ \cite{Meineri2020}.\footnote{Reflection positivity is only sufficient to show that \eqref{eq:bdy} is bounded in the limit \cite{Billo:2016cpy}, while the smoothness condition implies the even stronger bound; $\abs{c_1-c_2}^{2/3}/{\rm N}_\mathrm{D} $ stays finite in the limit ${\rm N}_{\rm D}\to 0$, see the Supplemental Material.}
We obtain the OPE
\begin{equation}\label{eq:DeltaD}
\begin{aligned}
&\pa{T(z)+\overline{T}(z)}\hat{\mathrm{D}}(w) \\
&=\fr{2}{(z-w)^2}\hat{\mathrm{D}}(w)+\fr{1}{(z-w)}
\del \hat{\mathrm{D}}(w)+ O(1).
\end{aligned}
\end{equation}
Hence, $\lim_{{\rm N}_{\rm D} \to 0}\hat{\mathrm{D}}$ is indeed an interface primary of weight $2$. It is worth emphasizing that this expression shows that $\hat{\mathrm{D}}(w)$ does not decouple from the theory.

Given that $\hat{\mathrm{D}}$ is an interface operator of conformal dimension 2,  
this constrains the possible bulk local operators that can emerge in the ${\rm N}_{\rm D} \to 0$ limit.  
In the following, we examine each candidate and show that the only viable possibility is a primary operator with conformal dimensions $(1,1)$.
We broadly distinguish two cases:  
a highest weight state under the conformal symmetry $\mathrm{Vir} \otimes \overline{\mathrm{Vir}}$ ({\it i.e.} a primary state),  
and those that can be obtained by acting with the stress tensor modes $L_{-n}$ on primary states ({\it i.e.} descendant states).

\noindent {\bf Primary fields:}
\smallskip
Suppose we have a bulk primary field $\phi (z, \bar z)$ of conformal weights 
$(h, \bar h)$. 
The OPE between a primary operator and the stress tensor is given by
\begin{equation}\label{eq:TD}
\begin{aligned}
T(z)\phi(w,\bar{w}) = \fr{ h }{(z-w)^2}\phi(w,\bar{w})+\fr{\del \phi(w,\bar{w})  }{(z-w)}+ O(1), \\
\overline{T}(\bar{z})\phi(w,\bar{w}) = \fr{ \bar{h} }{(\bar{z}-\bar{w})^2}\phi(w,\bar{w})+\fr{ \bar{\del} \phi(w,\bar{w})}{(\bar{z}-\bar{w})}+ O(1).
\end{aligned}
\end{equation}
Restricting $z$ and $w$ to the real axis and taking the difference between the two expressions, we obtain
\begin{equation}
    \pa{T(z) - \overline{T}(z)}\phi(w)
    = \frac{h - \bar h}{(z - w)^2}\phi(w)
    + \frac{\partial - \bar \partial}{z-w} \phi(w)
    + O(1).
\end{equation}
By symmetrizing with respect to $z$ and $w$, we obtain
\begin{equation}\label{symmetricterm}
\begin{aligned}
 & \pa{T(z) - \overline{T}(z)}\phi(w)+   \pa{T(w) - \overline{T}(w)}\phi(z) \\
&= \frac{h-\bar h }{(z-w)^2}(\phi(z) + \phi(w)) + O(1).   
\end{aligned}
\end{equation}
Recalling that $\phi$ is constructed from $T_1(w) - T_2(w)$,  
this expression implies that the OPE between $T_1(z) - \overline{T}_1(z)$ and $T_1(w) - T_2(w)$ contains symmetric singular terms, unless $h - \bar{h} = 0$.
On the other hand, the interface boundary conditions on the energy-momentum tensor imply that the singular parts of the OPE are anti-symmetric in $z$ and $w$,
\begin{equation}\label{eq:TD2}
\begin{aligned}
&\pa{T_1(z)- \overline{T}_1(z)}\pa{T_1(w) - T_2(w)} \\
&= T_1(z)\overline{T}_1(w)-\overline{T}_1(z) T_1(w) + O(1),
\end{aligned}
\end{equation}
where we used the fact that $T_1(z) \overline{T}_2( w)$ and  $\overline{T}_1(z) T_2(w)$ do not have singularities at $z=w$.
Consequently, we must have $h = \bar h$.
Since $h + \bar h = 2$, the bulk primary field $\phi$ that can appear in $\hat{\mathrm{D}}$ must be of conformal weights $(1,1)$, {\it i.e.}, marginal.

\noindent {\bf Descendant fields, $L_{-2}\ |0\rangle$ or $\overline{L}_{-2} \ |0\rangle$:}
The local operator corresponding to $L_{-2} |0\rangle$ is the energy-momentum tensor. It cannot appear by itself in $\hat{\mathrm{D}}(z)$, since the OPE of $T(z) + \overline{T}(\bar z)$ with $T(z)$ contains a $(z-w)^{-4}$ singularity due to the conformal anomaly, which contradicts the expression given in \eqref{eq:DeltaD}. However, if we instead consider the combination $\bar{c}T(z) - c\overline{T}(z)$, such a higher-order pole may cancel on the real axis. 

Nevertheless, this possibility can be excluded as follows.  
On the one hand, we have
\begin{equation}
    \pa{T(z) - \overline{T}(z)}\pa{\bar{c}T(w) -  c\overline{T}(w)}
    = \frac{c\bar{c}}{(z-w)^4} + \cdots,
\label{higherordersingularity}    
\end{equation}
where $z$ and $w$ are restricted to the real axis.  
On the other hand, as shown in \eqref{eq:TD2}, the OPE singularities between $T(z) - \overline{T}(z)$ and $\hat{\mathrm{D}}(w)$  
must be anti-symmetric under the exchange of $z$ and $w$.  
Since there are no other operators that can produce a $(z-w)^{-4}$ singularity to cancel the conformal anomaly term in \eqref{higherordersingularity},  
we conclude that $\bar{c}T(z) - c\overline{T}(z)$ cannot appear in $\hat{\mathrm{D}}(z)$.

\noindent {\bf Descendant fields, $L_{-1}\ |1, 0 \rangle$ or $\overline{L}_{-1} \ |0, 1 \rangle$:}
In this case, the CFT must possess either a holomorphic current $J$ or an anti-holomorphic current $\overline{J}$.
Suppose that $\hat{\mathrm{D}}(z)$ contains
a linear combination of $\partial J$ and $\bar \partial \overline{J}$.
Since we have
\begin{equation}
    \begin{aligned}
     &\pa{T(z) + \overline{T}(z) } \partial J(w)\\
     &=\frac{2  J(w) }{(z-w)^3}
     + \frac{2 \partial J(w) }{(z-w)^2}      + \frac{\partial^2 J(w) }{z-w} + O(1), \\
     & \pa{T(z) + \overline{T}(z) } 
      \bar \partial \overline{J}(w)\\
      &=  
     \frac{2\overline{J}(w)}{(z-w)^3} 
     + \frac{2  \bar   \partial \overline{J}(w)}{(z-w)^2} 
      + \frac{\bar \partial^2 \overline{J}(w) }{z-w} + O(1),
    \end{aligned}
\end{equation}
in order to satisfy  \eqref{eq:DeltaD}, we need to consider the combination $(\partial J- \alpha \bar \partial \overline{J})$ for some constant $\alpha$ and impose the totally reflective boundary condition $J(z) = \alpha \overline{J}(\bar z)$.
However, $J(z) = \alpha \overline{J}(z)$ on the real axis also implies $(\partial J - \alpha \bar \partial \overline{J})$ vanishes there. 
We thus conclude that $\hat{\mathrm{D}}(z)$ cannot contain either $\partial J$ or $\bar \partial \overline{J}$.

\noindent {\bf Descendant fields, $L_{-1}\ |1/2, 1/2 \rangle$ or $\overline{L}_{-1} \ |1/2, 1/2 \rangle$:}
Suppose there exists a bulk primary field $\rho$ of conformal weight $(1/2, 1/2)$.
Then, we have
\begin{equation}
\begin{aligned}
    \pa{T(z) + \overline{T}(z)} \rho(w, \bar w) 
    = \left(\frac{1/2}{(z-w)^2} + \frac{1/2}{(z - \bar w)^2} \right)\rho(w, \bar{w}) \\
    + \left( \frac{\partial }{z-w} 
    + \frac{\bar \partial}{z-\bar w} \right)
    \rho(w, \bar w)
    + O(1),
\end{aligned}
\end{equation}
for $z$ on the real axis.
It then follows
\begin{equation}
    \begin{aligned}
     &\pa{T(z) + \overline{T}(z) } \partial 
     \rho (w, \bar w)\\
      &=  
     \frac{ \rho(w, \bar w)}{(z-w)^3}
     + \frac{2\partial \rho(w, \bar w)}{(z-w)^2} 
     + \frac{(\partial+ \bar \partial) \partial \rho(w, \bar w) }{z-w} + O(1), \\
      &\pa{T(z) + \overline{T}(z) } \bar \partial \rho(w, \bar w) \\
      &=  
     \frac{\rho(w, \bar w)}{(z-w)^3} 
     + \frac{2\bar \partial \rho(w, \bar w)}{(z-w)^2} 
     + \frac{ (\partial + \bar \partial)\bar \partial \rho(w, \bar w)}{z-w} + O(1).
    \end{aligned}
\end{equation}
It implies that the combination $(\partial - \bar \partial) \rho$ satisfies \eqref{eq:DeltaD}. 

To summarize, we have shown that the conformal symmetry alone guarantees that the limit of $\hat{\rm D}$ is a linear combination 
\begin{align}\label{eq:diplacementfinal}
\lim_{{\rm N}_{\rm D}\to 0}\hat{\rm D}=\phi+(\partial -\bar{\partial})\rho\,,
\end{align}
where $\phi$ and $\rho$ are dimension $(1,1)$ and $(\frac{1}{2},\frac{1}{2}) $ primaries respectively.
Both $\phi$ and $\rho$ must be neutral under any global symmetry that preserves the interface since $\hat D$ is.
Since we assumed that the theory does not contain neutral $(\frac{1}{2},\frac{1}{2})$ primaries, this establishes the existence of a marginal operator (since $\rho=0$ and therefore $\phi$ is necessarily present). We will elaborate more on this point in the next section.

Let us now verify that the three-point function of $\phi$ vanishes, as expected from exact marginality. 
Given a pair of quasiprimary operators $O_1$ and $O_2$, with the same spin $h_1-\bar{h}_1=h_2-\bar{h}_2$, on either side of any conformal interface, their correlators with and without the displacement are constrained as follows \cite{Gliozzi:2015qsa}:
\begin{align}    &\braket{O_1(z,\bar{z})O_2(\bar{z},z)}=\frac{a_{12}}{\pa{2\mathrm{Im} z}^{\Delta_1+\Delta_2}} \label{OLOR} \\
&\braket{O_1(z,\bar{z})O_2(\bar{z},z)\hat{\rm D}(w)}=
    \frac{\pa{\Delta_1-\Delta_2}a_{12}/ \sqrt{\mathrm{N}_{\mathrm{D}}}}{\pa{2\mathrm{Im} z}^{\Delta_1+\Delta_2-2}|z-w|^4}~, \label{OOD}
\end{align}
where $\Delta=h+\bar{h}$, $w \in \mathbb{R}$, $a_{12}$ is a constant which, at $\mathrm{N}_{\mathrm{D}}=0$, is fixed by normalization, {\it i.e.}, $a_{12}=1$, and it should be noted that $O_1$ and $O_2$ are placed at symmetric positions with respect to the interface.   Having established the existence of an operator of weight 2 at a generic point of the conformal manifold, we select $O_1=O_2=\phi$, so that the r.h.s. of \eqref{OOD} vanishes identically for any value of $\mathrm{N}_{\mathrm{D}}$, while \eqref{OLOR} does not, and in particular $\lim_{\mathrm{N}_{\mathrm{D}}\to 0}a_{12}=1$.
Since the r.h.s. of \eqref{OOD} reduces to the three-point function of $\phi$ at $\mathrm{N}_{\mathrm{D}}=0$, we conclude that 
\begin{equation}
    \langle \phi(z) \phi(w)\phi(u) \rangle = 0\,,
    \label{marginality}
\end{equation}
establishing exact marginality of $\phi$ at the leading order.

More generally, \eqref{OLOR} and \eqref{OOD} allow us to derive the expected relation between the three-point functions involving $\phi$ and the leading order change $\delta \Delta$ of the scaling dimensions along the corresponding direction in the conformal manifold. Indeed, considering a generic quasiprimary $O$ and repeating the reasoning above, one obtains
\begin{equation}
    \delta \Delta_O = \sqrt{\mathrm{N}_{\mathrm{D}}}\, c_{OO\phi} + o\pa{\sqrt{\mathrm{N}_{\mathrm{D}}}}~.
\end{equation}

\section{Interface Conformal Manifold}

Why did we need to assume the absence of neutral $(\frac{1}{2},\frac{1}{2})$ primaries? Let us first clarify what $\rho$ actually represents in \eqref{eq:diplacementfinal}. If a CFT contains a weight one scalar operator, one can produce a line defect by integrating it with an infinitesimal coefficient. If, furthermore, $c_{\rho\rho\rho}=0$, together with the required higher order conditions, the line defect is part of a conformal manifold. As a concrete example, the Ising CFT admits a defect conformal manifold generated by the following interface exactly marginal deformation \cite{Oshikawa1997},
\begin{equation}
S_{\textup{defect}} = \gamma \int dx \  \varepsilon.
\end{equation}
where $\varepsilon$ is the energy density operator with conformal dimensions $(1/2, 1/2)$.
The displacement operator is:
\begin{equation}
\mathrm{D} = \fr{\gamma}{2} (\del-\bar{\del}) \varepsilon.
\end{equation}
This is precisely of the same form as the second term in (\ref{eq:diplacementfinal}). Vice versa, given a family of interfaces connected to the identity, whose displacement obeys (\ref{eq:diplacementfinal}) with $\phi=0$, one can prove that $c_{\rho\rho\rho}=0$---see the Supplemental Material. Furthermore, it is easy to check that, if $\lim_{\mathrm{N}_{\mathrm{D}}\to 0} \hat{\rm D}=(\partial-\bar{\partial})\rho$, the l.h.s. of \eqref{OOD} vanishes in the limit, therefore no scaling dimension changes at leading order in $\mathrm{N}_{\mathrm{D}}.$
More generally, symmetry-neutral exactly marginal deformation on the interface can arise from the breaking of  a {\it phantom symmetry} \cite{Antinucci2025}: a symmetry that emerges when one interprets the interface between $\mathrm{CFT}_1$ and $\mathrm{CFT}_2$ as a boundary in $\mathrm{CFT}_1 \otimes \overline{\mathrm{CFT}_2}$ using the folding trick. In such cases, a {\it phantom current} with conformal dimension $(1,0)$ in the folded theory induces an interface exactly marginal operator with dimension $(1/2,1/2)$ in the unfolded theory. This is precisely the case in the 
example of the Ising CFT, where the phantom current is given by $j = \psi_1 \psi_2$.
Here, $\psi_{1/2}$ represents the fermion field on the upper/bottom side of the interface,  
and the energy density operator in the Ising CFT can be expressed in terms of these fermions as 
$\varepsilon_{1/2} = i \psi_{1/2} \bar{\psi}_{1/2}$.

The explicit examples discussed above describe interfaces between identical CFTs, rather than interfaces connecting different points on a conformal manifold as considered in the main text. However, our leading-order analysis does not rule out the possibility that an interface could connect theories at distinct points on a conformal manifold while still obeying $\lim_{\mathrm{N}_{\mathrm{D}}\to 0} \hat{\rm D}=(\partial-\bar{\partial})\rho$. Consider, for instance, the case where ${\rm D} \sim \tilde{\phi} + (\partial-\bar{\partial})\rho$ with $\braket{\tilde{\phi}\tilde{\phi}} =O(\mathrm{N}_{\mathrm{D}}^{1+\epsilon})$ with $\epsilon > 0$. Equation \eqref{eq:diplacementfinal} would be obeyed with $\phi=0$. This example shows that additional input is needed to treat conformal manifolds containing protected dimension $1$ operators within our framework.\footnote{Of course, if a scalar has dimension 1 only on some submanifold of the conformal manifold, our analysis works by considering points in the complement of said submanifold.}

\section{Examples}
Let us illustrate our results with concrete examples:  
the compact free boson, whose action is given by
\begin{equation}
S=\fr{1}{4\pi} \int d^2z \  \del X \bar{\del} X, \ \ \ \ \ X\sim X+2\pi R,
\end{equation}
where $R$ represents the compactification radius.
The marginal operator in this CFT is $\phi = \del X \bar{\del} X$.
The exactly marginal deformation leads to the shift in the compactification radius, {\it i.e.} $R \to R'\equiv R\ex{\pi \delta\lambda}$.
The interface with a CFT deformed by $\delta \lambda$ can be described by the following gluing matrix \cite{Bachas2002},
\begin{equation}
  \left(
    \begin{array}{c}
       \del X_1   \\
       \bar{\del}X_2  \\
    \end{array}
  \right)
  =
  S
    \left(
    \begin{array}{c}
       \bar{\del}X_1   \\
       \del X_2  \\
    \end{array}
  \right),
  \ \ \ \ \ \ \ 
  S = 
    -\left(
    \begin{array}{cc}
     \cos 2\theta  & -\sin 2\theta   \\
      \sin 2\theta &  \cos 2\theta  \\
    \end{array}
  \right),
\end{equation}
where we define $\tan \theta = \ex{\pi \delta\lambda}$.
This gluing condition allows us to express the stress tensor on the bottom side of the interface in terms of the free boson fields on the upper side,
\begin{equation}
\begin{aligned}
T_2&=\fr{1}{2} \del X_2 \del X_2 = \fr{1}{2}\del X_1 \del X_1 - \pi \delta\lambda \del X_1 \bar{\del} X_1,  \\
\overline{T}_2&=\fr{1}{2} \bar{\del} X_2 \bar{\del} X_2 = \fr{1}{2} \bar{\del} X_1 \bar{\del} X_1 - \pi \delta\lambda \del X_1 \bar{\del} X_1. 
\end{aligned}
\end{equation}
This expression leads to the following displacement operator,
\begin{equation}
\lim_{\delta\lambda \to 0} \hat{\mathrm{D}} = \del X \bar{\del} X.
\end{equation}
This coincides precisely with the exactly marginal operator $\phi = \del X \bar{\del} X$.

Another example is the supersymmetric sigma-model whose target space is a Calabi-Yau manifold. For each exactly marginal deformation, there is a primary field $\rho$ of dimensions $(1/2, 1/2)$ and
an exactly marginal operator $\phi$, which is a superconformal descendant of $\rho$. However, since $\rho$ carries non-zero $R$ charges with respect to the ${\cal N} =(2,2)$ superconformal algebra, it cannot appear on the right-hand side of \eqref{eq:diplacementfinal}.
For a generic Calabi-Yau manifold, there are no neutral $(1/2, 1/2)$ primaries. Therefore, the limit of $\hat D$ can be identified with the exactly marginal operator $\phi$.

\section{Comments}

In the introduction, we noted the interesting analogy between the absence of free parameters and the absence of global symmetry in AdS gravitational theories, both of which are consequences of the splittability of symmetry generators. The absence of free parameters has been quantified as Distance Conjectures 1 and 2
\cite{Ooguri:2006in} and formulated in AdS \cite{Perlmutter:2020buo}, and some of these conjectures have been proven \cite{Baume:2023msm,Ooguri:2024ofs}. The absence of global symmetry has been quantified as the Weak Gravity Conjecture \cite{Arkani-Hamed:2006emk} and formulated in AdS \cite{Nakayama:2015hga,Harlow:2015lma,Harlow:2018tng,Aharony:2021mpc}, and some of these have been proven \cite{Sharon:2023drx,Aharony:2023ike}. Since both are formulated in terms of inequalities, it would be interesting to determine whether they are analogous and can be understood within a single framework.

\section*{Acknowledgments}
We thank Tom Banks, Nathan Benjamin, Davide Gaiotto, Daniel Harlow, Joao Penedones, Ashoke Sen, Mark Van Raamsdonk, Yifan Wang, Yunqin Zheng and Sasha Zhiboedov for discussions.
YK and HO are supported in part by the Walter Burke Institute for Theoretical Physics at Caltech and by the U.S. Department of Energy, Office of Science, Office of High Energy Physics, under Award Number DE-SC0011632.
YK has also been supported in part by the Brinson Prize Fellowship at Caltech, by the INAMORI Frontier Program at Kyushu University, and by JSPS KAKENHI Grant Number 23K20046.
HO is also supported in part by the  Leinweber Forum for Theoretical Physics at Caltech, 
the Simons Investigator Award (MP-SIP-00005259), and
JSPS Grants-in-Aid for Scientific Research 23K03379. 
His work was performed in part at
the Kavli Institute for the Physics and Mathematics of the Universe at the University of Tokyo, which is supported by the World Premier International Research Center Initiative, MEXT, Japan,  at the Kavli Institute for Theoretical Physics (KITP) at the University of California, Santa Barbara, which is supported by NSF grant PHY-2309135, at the Aspen Center for Physics, which is supported by NSF grant PHY-1607611, and Pioneering Science Promotion Division of RIKEN. MM is supported by the Italian Ministry of University and Research (MUR) under the FIS grant BootBeyond (CUP: D53C24005470001), and by the INFN ``Iniziativa Specifica'' ST\&FI. SK and YK thank Yukawa Institute for Theoretical Physics and the workshop ``Progress on Theoretical Bootstrap" for hospitality during the completion of this work.

\bibliographystyle{JHEP}
\bibliography{main.bib}

\newpage

\onecolumngrid

\begin{center}
    \LARGE \textbf{Supplemental Material}
\end{center}
In this Supplemental Material, we derive a few technical results on the limit of three-point functions which complement the discussion in the main text.
\subsection{Three-Point Function of $\hat{\rm D}$.}
We first compute the three-point function of the normalized displacement operator $\hat{\rm D}\equiv {\rm D}/\sqrt{{\rm N}_{\rm D}}$. For this purpose, we consider the OPE of $T_1-T_2$,
\begin{align}
\begin{aligned}
&(T_1(x) - T_2(x)) (T_1(y) - T_2(y)) (T_1(z) - T_2(z))\\
&=(\bar{T}_1(x) - \bar{T}_2(x)) (T_1(y) - T_2(y)) (T_1(z) - T_2(z)) \\
&= \bar{T}_1(x)  T_1(y) T_1(z) - \bar{T}_2(x)  T_2(y) T_2(z) \\
&= \fr{c_1\bar{T}_1(x)-c_2\bar{T}_2(x)  }{2(y-z)^4}
+\pa{\fr{2}{(y-z)^2} + \fr{\partial_z}{(y-z)} } \pa{ \bar{T}_1(x)T_1(z) - \bar{T}_2(x) T_2(z)  }\\
&= \fr{c_1\bar{T}_1(x)-c_2\bar{T}_2(x)  }{2(y-z)^4}
+\pa{\fr{2}{(y-z)^2} + \fr{\partial_z}{(y-z)} } \pa{ T_1(x)T_1(z) - T_2(x) T_2(z)  }\\
&= \fr{c_1\bar{T}_1(x)-c_2\bar{T}_2(x)  }{2(y-z)^4} 
 +\pa{\fr{2}{(y-z)^2} + \fr{\partial_z}{(y-z)} }  \left[\frac{c_1-c_2}{2(x-z)^4}+\pa{\fr{2}{(x-z)^2} + \fr{\partial_z}{(x-z)} }   \pa{ T_1(z) -  T_2(z)  }\right].
\end{aligned}
\end{align}
Since the one-point functions on the interface vanish because of the conformal symmetry, we conclude that
\begin{align}
\langle (T_1(x) - T_2(x)) (T_1(y) - T_2(y)) (T_1(z) - T_2(z))\rangle=\frac{c_1-c_2}{(y-z)^2(x-z)^{4}}\left(1+O((y-z))+O((x-z))\right)
\end{align}
Dividing each operator by the normalization $\sqrt{{\rm N}_{\rm D}}$ and comparing the result with the expected behavior of the three-point function, $\sim (x-y)^{-2}(y-z)^{-2}(z-w)^{-2}$, we can read off the structure constant
\begin{align}
c_{\hat{\rm D}\hat{\rm D}\hat{\rm D}}=\frac{c_1-c_2}{{\rm N}_{\rm D}^{3/2}}\,.
\end{align}
Thus, the smoothness assumption implies that $\abs{c_1-c_2}^{2/3}/{\rm N}_\mathrm{D}$ is finite in the limit ${\rm N}_{\rm D}\to 0$.

\subsection{Interface conformal manifolds and $c_{\rho\rho\rho}$}

Suppose that a pair of CFTs (perhaps identical) admits a family of conformal interfaces which includes the trivial one, such that
\begin{equation}
\lim_{{\rm N}_{\rm D}\to0} \hat{\rm D} = i(\partial-\bar{\partial}) \rho~,
\label{Disprho}
\end{equation}
where $\rho$ is a dimension one scalar as in the main text. Here we added an $i$ to make the displacement real for real $\rho$.
In this appendix, we prove that the three-point function of $\rho$ vanishes. While the argument starting with equations (\ref{OLOR},\ref{OOD}) can be repeated with $O_1=O_2=\rho$, this is to no avail because $\braket{\rho(z,\bar{z})\rho(\bar{z},z)(\partial-\bar{\partial}) \rho(w,w)}$ vanishes kinematically. Instead, we will derive the result as a consequence of conformality of the interface, together with the Ward identities of the displacement. In the rest of this appendix, we use the real coordinates $x=\Re z,\,y=\Im z$, and write $\rho(x,y)$ slightly redefining our notation.

Consider the following equation:
\begin{equation}
    \braket{\rho(0,y_0+\delta y)}=
    \frac{a_\rho({\rm N}_\Disp)}{2y_0} \left(1-\frac{\delta y}{y_0}+\left(\frac{\delta y}{y_0}\right)^2+ O\left(\delta y^3\right)\right)~.
    \label{taylor1point}
\end{equation}
The dependence on $\delta y$ must be reproduced by the following deformation of the defect action:
\begin{equation}
    \braket{\rho(0,y_0+\delta y)}=\frac{\left\langle \rho(0,y_0)\, e^{-\delta S}\right\rangle}{\left\langle  e^{-\delta S}\right\rangle}~, \qquad
    \delta S = \int\! dx\, \left(\delta y \Disp(x)+O(\delta y^2)\right)~.
    \label{dispAction1point}
\end{equation}
The leading order expansion of \eqref{dispAction1point} leads to the usual Ward identity
\begin{equation}
   \frac{\del}{\del y} \braket{ \rho(0,y)}=-\int\! dx\,\left\langle\rho(0,y) \Disp(x)\right\rangle~.
   \label{WardDisp}
\end{equation}
Equation \eqref{WardDisp} implies that $a_{\rho}$ starts at order $\sqrt{{\rm N}_{\rm D}}$, and furthermore $\rho$ is the only operators whose one-point function has this property. 

Equation \eqref{taylor1point} showcases infinitely many terms of order $\sqrt{{\rm N}_{\rm D}}\, (\delta y)^n$, with $n$ integer, while the leading term in $\delta S$ \eqref{dispAction1point} would lead to an expansion in powers of $({\rm N}_{\rm D})^{m/2}\, (\delta y)^n$, with $m \geq n$. The only possible corrections to $\delta S$ with the desired properties at order $\sqrt{{\rm N}_{\rm D}}$ involve descendants of $\rho(x,y)$.\footnote{Of course, if we were to assume that the line defect has action $S_\textup{defect}=\int dx\, \rho(x,0)$, we could immediately conclude that $\delta S$ in \eqref{dispAction1point} contains the full Taylor expansion of $\rho(x,\delta y)$ around $\delta y=0$. While our only assumption is \eqref{Disprho}, it is useful to keep in mind this more restricted scenario.}
It will be sufficient to stop at order $(\delta y)^2$:
\begin{equation}
    \delta S = \int\! dx\, \left(\delta y\, \Disp(x)-g \delta y^2 \left(\partial_y^2\rho(x,0)+\kappa\, \partial_x^2\rho(x,0) \right)+O(\delta y^3)\right)~.
    \label{dispAction}
\end{equation}
The defect descendant $\partial_x^2\rho$ has no effect, and it is included for later convenience. Instead, we have not included the dimension one operator $\rho$ in the action, despite it being more relevant. The reason is that, by dimensional analysis, it cannot appear with a positive power of $\delta y$ in the continuum limit, hence it is not needed to reproduce eq. \eqref{taylor1point} and should be tuned to zero.

We can now compute the order $(\delta y)^2$ of equation \eqref{dispAction1point}, in a small-${\rm N}_{\rm D}$ expansion. With foresight, we define
\begin{equation}
    g = g^{(1)} \sqrt{{\rm N}_{\rm D}} + g^{(2)} {\rm N}_{\rm D}+\dots,
\end{equation}
and obtain
\begin{multline}
\frac{1}{2}\frac{\partial^2}{\partial y^2} \braket{\rho(0,y)} = g^{(1)} \sqrt{{\rm N}_{\rm D}}  \int\! dx\,\braket{\rho(0,y)\,\partial_y^2 \rho(x,0)}_0 \\
+ {\rm N}_{\rm D} \left[ \int\! dx_1\! \int\! dx_2 \braket{\rho(0,y) \partial_y \rho(x_1,0)\partial_y \rho(x_2,0)}_{0}
+ g^{(2)}   \int\! dx\,\braket{\rho(0,y)\,\left(\partial_y^2 \rho(x,0)+\kappa\, \partial_x^2\rho(x,0)\right)}_0 \right]~.
\label{Odysquared}
\end{multline}
In this expression, the subscript 0 denotes expectation values taken in the absence of the defect. 
Crucially, the double integral contains divergences, which must be regulated and renormalized. For concreteness, we choose a short-distance cutoff $a$ with dimension of length. Then, a power-law divergence $\propto a^{-2}$ should be cured by a counterterm proportional to $\rho(x,0)$, which we disregard as discussed above. Instead, a logarithmic divergence generates the coupling $g^{(2)}$ along the RG flow, and we focus on the latter. In detail, we get
\begin{equation}
 \int\! dx_1\! \int\! dx_2 \braket{\rho(0,y) \partial_y \rho(x_1,0)\partial_y \rho(x_2,0)}_{0} \sim \dots +
 \left(\log \frac{\ell}{a}\right) c_{\rho\rho\rho}\!\int\! dx\, \frac{4 x^2+48 y^2}{(x^2+4 y^2)^3}+ \textup{finite}~,
 \label{divergenceCrrr}
\end{equation}
where $\ell$ is an IR cutoff, which would cancel out against the finite part of the integral. The logarithmic divergence can be canceled by a counterterm proportional to $g^{(2)}$:
\begin{equation}
    g^{(2)}= g^{(2)}_2(\mu) + \frac{5}{16}\, c_{\rho\rho\rho} \,\log a \mu ~,
    \quad\qquad \kappa = \frac{3}{5}~.
    \label{g2ct}
\end{equation}
The coupling to $\partial^2_x \rho$ in the action \eqref{dispAction} allows to cancel the divergence locally, at the level of the integrand, rather than upon integration. In \eqref{g2ct} we introduced the RG scale $\mu$, which, as usual, allows to track the effect on observables of rescaling the cutoff. Since the bare coupling $g^{(2)}$ does not depend on $\mu$, one gets the obvious beta function
\begin{equation}
    \mu \frac{\partial g_2^{(2)}}{\partial\mu}=\beta (g_2^{(2)}) = -\frac{5}{16}\, c_{\rho\rho\rho}~.
    \label{betag2}
\end{equation}
Equation \eqref{betag2} shows that the corresponding coupling cannot be fine-tuned if the three-point function of $\rho$ does not vanish: it is always generated and evolves along the RG flow. This violates the hypothesis that the interface is conformal for any value of ${\rm N}_{\rm D}$ sufficiently small, hence we conclude
\begin{equation}
    c_{\rho\rho\rho}=0~.
    \label{crrr0}
\end{equation}
With little more work, one can find the explicit anomalous dependence of the one-point function of $\rho$ on $\log y$, proportional to $c_{\rho\rho\rho}$.

\end{document}